# Elongation of discotic liquid crystal strands and lubricant effects

Surjya Sarathi Bhattacharyya and Yves Galerne*

*Institut de Physique et Chimie des Matériaux de Strasbourg*

*UMR 7504 (CNRS – Université de Strasbourg)*

*23 rue du Lœss, 67034 Strasbourg, France*

## Abstract

After a short review on the physics of pulled threads and their mechanical properties, the paper reports and discusses on the strand elongation of disordered columnar phases, hexagonal or lamello-columnar, of small molecules or polymers. The mechanical properties appear to be relevant to the length of the columns of molecules compared to the thread length, instead of the usual correlation length. When short, the column entanglement being taken into account, the strand exhibits rather fluid properties that may even look like nematic at a macroscopic scale. Then, the Plateau-Rayleigh instability soon breaks the thread. However, the hydrodynamic objects being the columns instead of the molecules, the viscosity is anomalously large. The observations show that the strands of columnar phases are made of filaments, or fibrils, that indeed are bundles of columns of molecules. They both explain the grooves and rings observed on the antenna or bamboo-like strand profiles. On pulling a strand, the elongation stress eventually exceeds the plasticity threshold, thus breaking columns and filaments. Cracks, more exactly, giant dislocations are thus formed. They change the strand thickness by steps of different birefringence colours. Interestingly, adding a solute may drastically change the effective viscosity of the columnar phase and its mechanical properties. Some solutes as alcanes, exhibit lubricant and detangling properties, while others as triphenylene, are quite anti-lubricant.

* yves.galerne@ipcms.unistra.fr

## 1. Introduction

Known for a long time, the discotic liquid crystals (LCs) have not been studied so intensively as the calamitic LCs up to now, though indeed they have a real potential for practical applications.[1 – 4] The reason is essentially that they are not easy to orient in uniform samples, and that consequently, the experiments are difficult to perform. Somehow, the situation is reminiscent to the case of the chiral smectic C* phase, whose ferroelectric properties were so promising for display applications 40 years ago.[5] Unfortunately, this phase turned out to be extremely difficult to orient in the so-called bookshelf geometry where promising applications were predicted. The method is therefore actually abandoned after years of active, but unsuccessful researches. The reason of a so difficult alignment originates from the 1D positional order of the smectic layers that adds up to the orientational order of LCs, and that forbids twist and bend distortions usually available in nematics. The consequence is that smectics organize in focal conics only.[3] The constraint is strong as the positional order of smectics is associated to a relatively stiff elastic constant $\sim kT / d^3$, where $d$ is the molecule diameter, $T$ the temperature and $k$ the Boltzmann constant.[2] This makes the phase unable to adapt easily to slightly irregular anchoring properties on the substrates that usually do not make any problem to the nematic phase. The columnar phases are even more difficult to orient uniformly, even when slowly decreasing the temperature from a uniformly oriented nematic sample – that may be obtained rather easily through the usual methods that are essentially based on surface treatments as for the calamitic LCs. As in the smectic case, the positional order of the molecules in the columnar phases needs to be controlled in addition to the orientational order of the nematic phase. In fact, the discotic columnar phases exhibits a 2D positional order, and are therefore closer to the crystal phase (3D positional order) than to the nematics. This basically explains that they are globally more difficult to orient than the smectics (1D). As discussed earlier by Kleman, they organize in developable domains.[6, 4]

Preparing free-standing films has been a successful method for realizing well-oriented smectic samples. [4] Perfectly oriented samples with the layers parallel to the free surfaces are obtained in this manner, and generally they are stable for several hours. The method has been extended to free-standing strands of discotic LCs.[7, 8] Essentially, three discotic LCs are available in the sense that these phases exhibit positional disorder along one direction at least, and therefore they may exhibit some of the properties of liquids. There is the nematic phase where the discotic molecules have no positional order in any of the 3 dimensions of space, though they are oriented parallel to one another except for thermal fluctuations. There are also the columnar phases where the molecules (again parallel to one another) are now gathered in closely packed columns. Two sub-classes exist then. The columns being parallel to one another, they may constitute a hexagonal 2D array if the molecules are perpendicular to the columns, or a rectangular array if they are tilted at an angle in the columns. In this latter case, the phase is called lamello-columnar. Somehow, these two columnar phases may be considered as 1D liquids, and 2D crystals. They are analogous to the smectic A, C, $C_A$, O phases, etc… that are obtained with calamitic molecules. These smectics are 2D liquids and 1D crystals, while let us recall it, smectic B, D, E, G, H phases, etc... are just 3D molecular crystals.

Different methods have been developed for preparing 1D nano or micro-objects of discotic LCs. They use different processes as vapour deposition, template synthesis, precipitation from solutions, or the elongation by drawing a thread with a needle from a drop, etc… [9] Here, we focus on this pulling method, and we discuss why so different behaviours are evidenced between the samples, some of them being easy to manipulate while others immediately break. In particular, we report on observations performed on different discotic compounds including a polymer, and that exhibit the three different phases, a nematic phase and two discotic disordered columnar phases, respectively hexagonal and lamello-columnar, where the molecules are not set at definite positions in the columns. Depending on the phase, but not only, the discotic LCs may be pulled in threads with more or less success. Typically, pulling threads in the nematic phase, that is indeed a 3D liquid, is really easy, but just lasts for a moment since the Plateau-Rayleigh instability will soon break them down. In comparison, the existence of a structure into the columnar phases drastically changes the mechanical properties of the material, and therefore changes the behaviour of the strands under elongation.

## 2. Nematic phase

Apart from its orientational properties, the nematic phase exhibits the usual properties of liquids. It is therefore easy to pull threads in the nematic phase. But, as there is no physical reason why the thread should keep the same diameter all along its length, a thinning process occurs after a while, namely, the Plateau-Rayleigh instability occurs and breaks the thread soon.

## 2. 1. Plateau-Rayleigh instability

The threads pulled in the nematic phase have typical lengths ~ 1 – 2 mm, and exhibit a conic shape with a diameter $2r$ that regularly decreases away from the menisci (Section 4. 2). Because of the Plateau-Rayleigh instability, their life time after the beginning of the pulling process is on the order of a couple of seconds. If one first neglects the effect of the elongation stress (Section 2. 2), the mechanism is quite simple ; the thread spontaneously thins and finally breaks down to reduce its surface energy, or more precisely, under the action of Laplace pressure. The excess energy is then partly converted in kinetic energy and partly dissipated by viscosity.[10] Rough estimates are sufficient for the discussion. The excess surface energy is on the order of $E_{surf} \sim \gamma \, r \, l$, where $r \sim 5$ μm and $l \sim$ a few times $r$, is the length of the thread involved in the pinching – i. e. $l$ corresponds to the wavelength of the instability – and $\gamma$ is the surface energy of the nematic phase. The exact value of $\gamma$ naturally depends on the nature of the molecules and on temperature. Nevertheless, based on measurements performed on resembling discotic molecules, we may estimate $\gamma \sim 10^{-2}$ J m$^{-2}$.[11] The excess surface energy $E_{surf}$ is therefore equal to the addition of the kinetic energy of the thread, the order of magnitude of which is given by $E_{kin} \sim \rho \, l \, r^2 \, \dot{r}^2$, and of the dissipation energy $E_{visc} \sim p_{visc} \times \pi \, r^2 \, l \times \tau$ . Here, $p_{visc} \sim \dfrac{1}{2} \eta \left( \dfrac{\dot{r}}{r} \right)^2$ is the power density of dissipation, $\dfrac{\dot{r}}{r} \sim \dfrac{1}{\tau}$ being the strain rate involved before the rupture of the thread, $\tau$ is the characteristic time for rupture, and $\eta$ the viscosity. The conservation of the thread energy therefore yields :

$$k_1 \, \rho \, r \, \dot{r}^2 + k_2 \, \eta \, \dot{r} + \gamma \sim 0 \, , \tag{1}$$

where $k_1$ and $k_2$ are numeric coefficients that we do not need here to express in detail. For a radius of the thread $r$ smaller than a characteristic radius $r* = \dfrac{\eta^2}{\gamma \rho}$, the first term of the equation (kinetic term), may be neglected. The process is then essentially viscous. We deduce the typical time after which a rupture occurs under the action of surface tension to be :

$$\tau_{visc} \sim \dfrac{\eta r}{\gamma} \, . \tag{2}$$

In the opposite case, for $r > r^*$, we may neglect dissipation. The time before rupture should then be on the order of $\tau_{\mathrm{kin}} \sim \left[\dfrac{\rho \, r^3}{\gamma}\right]^{1/2}$.

In order to discuss if the thread rupture occurs in the viscous or in the kinetic regime, we should know the order of magnitude of the viscosity of the discotic nematic phase. Unfortunately, the viscosity of LCs is rarely measured. Only few experimental data are available for small calamitic molecules, essentially in the *n*-alkylcyanobiphenyl series (nCB). Typically, for 5CB, we have $\eta_{5CB} \sim 5$ to $25 \times 10^{-3}$ Pa s.[12] For large molecules as discotics, the viscosity should be larger, but extrapolations are difficult to do. Conversely, we may use the experimental orders of magnitude $\tau \sim 0.5$s and $r \sim 5$ μm (Section 4. 2) and deduce from Eq. 1 that the kinetic term is completely negligible compared to $\gamma$. The rupture process relevant to the Plateau-Rayleigh instability is therefore essentially viscous. We may then deduce an order of magnitude of the viscosity from Eq. 2, to be $\eta \sim 1000$ Pa s. This simplified analysis could, however, underestimate the nematic viscosity since the contribution of the elongation stress to the breaking time is not taken into account here (see below).

## 2. 2. Elongation stress

When pulling a nematic thread with a force *F* at a velocity *v*, we produce an elongation stress that naturally competes with the Plateau-Rayleigh instability for thinning and breaking the thread. The elongation stress is essentially concentrated, as the strain rate, in a cylinder of length *l* around the place with minimum radius *r*. It may be estimated as

$$\frac{F}{\pi r^2} \sim \eta \frac{v}{l} \ . \tag{3}$$

Formally this elongation stress adds up to the Laplace pressure $\dfrac{\gamma}{r}$, for thinning and breaking the thread (Section 2. 1) , but its contribution is negligible if the aspect ratio of the nematic thread satisfies the condition

$$\frac{l}{r} > \frac{\eta v}{\gamma}. \tag{4}$$

With the above values $\eta \sim 1000$ Pa s and $\gamma \sim 10^{-2}$ J m$^{-2}$, and for an elongation velocity $v = 2\ \mu$m s$^{-1}$, the condition (4) becomes $\frac{l}{r} > 0.2$ which is obviously correct. So then, the Laplace pressure is the leading term, including in the final stage before the thread rupture. Clearly, the thinning process accelerates due to the Laplace pressure and the elongation stress that both concentrate precisely at the place where the thread is already the thinnest. Finally, this zone turns to a marked necking of length about equal to its diameter, i.e. $\frac{l}{r} \sim 2$,[13] and after an ultimate thinning of the thread down to a molecular size, the thread eventually breaks.

However, with a faster elongation velocity as $v = 500\ \mu$m s$^{-1}$ (Section 4. 2), or with other discotic liquid crystals that exhibit large viscosities (Section 4. 3. 1), we may observe that $\frac{\eta v}{\gamma}$ exceeds 2. The condition (4) is not fulfilled any more during the necking of the thread (where $l \sim 2r$). The elongation stress is now dominant, and according to Eq. 3, we may estimate it as $\sim \eta \frac{v}{2r}$. The rupture of the thread occurs when the elongation stress exceeds the cohesion stress $P_{coh}$. This limit may be estimated from the force per surface unit $P = \frac{A}{6\pi D^3}$ that two parts of the thread exert on each other at the minimum distance $D_{min} \sim 1$ nm, $A$ being the Hamaker constant between discotic molecules when they are parallel to each other.[14] This force $P$ when integrated to infinity yields the cohesion energy $U_{coh}$ of the material per surface unit. Therefore, on noticing that $U_{coh} = 2\gamma_{\perp}$, where $\gamma_{\perp}$ is the energy per surface unit perpendicular to the columns, and considering that $\gamma_{\perp} \sim \gamma$ [though the core part of the discotic molecules exerts a larger London interaction than the peripheral alkyl chains], we may estimate $P_{coh} \sim \frac{4\gamma}{D_{min}}$, i.e. $P_{coh} \sim 4 \times 10^7$ Pa. So, the thread breaks when the elongation stress $\eta \frac{v}{2r}$ exceeds $P_{coh}$, i. e. when its diameter gets below

$$2r \sim \frac{\eta v}{P_{coh}} \ .$$

(5)

With $\eta \sim 10^6$ Pa s (see for instance, Section 4. 3. 1) and for elongation velocities $v \sim 500$ μm s$^{-1}$, the rupture due to the elongation stress should occur for diameters $2r \sim 10$ μm. This discontinuity represents a marked difference from the rupture relevant to the Plateau-Rayleigh instability, which arises for smaller viscosities and slower (possibly null) pulling velocities, and that are characterized by thread radii continuously decreasing until a molecular size.

## 3. Disordered columnar phases

In contrast to the nematic phase, the disordered hexagonal columnar and disordered lamello-columnar phases, are not 3D liquids. They exhibit an internal structure that gives them basically different mechanical properties. However, the mechanical properties of their internal structure are not so well-cut. The molecules may jump rather easily from a column to a neighbouring one, and more importantly, the columns that they build have a finite length. Both properties may give some fluidity to the disordered columnar phases.

## 3. 1. Point defects in disordered columnar threads

Contrary to smectics that have been intensively studied, the disordered columnar discotics are rather poorly documented. However, there is a clear analogy between the smectic and columnar discotic phases since they exhibit 1D and 2D positional orders, respectively.[4] So, roughly speaking, the 1D problem of pulling a strand is akin to stretching a 2D smectic film on a frame, or to squeezing a smectic sample between plates, the smectic layers replacing the columns of molecules. Clearly, the analogy with the smectic phases allows us to extrapolate most of the results in smectics to the discotic columnar phases.[15] In both cases, when pulling a film or a thread, defects are created and they eventually migrate in the elongation direction. However, there are differences between both the problems precisely because of their different dimensionality. For instance, breaking a column of molecules in a columnar phase, or nucleating a new column, just needs to overcome an energy barrier of the order of the interaction energy between two molecules. In smectics, adding or suppressing a smectic layer needs to create a single-layer dislocation loop that costs energy until its size exceeds some critical radius.[16] So, the energy necessary to break a column, $E_b$ , in a columnar system is on the order of the interaction energy between two neighbouring molecules in a column, that is smaller by one order of magnitude at least than for breaking layers in smectic systems. This energy is essentially of the London type. It is generally worth a few $kT$ , and may simply be paid by the thermal energy with a Boltzmann probability $\sim \exp - \dfrac{E_b}{kT}$ . Consequently, the density of point defects is finite in the columnar phases, as the length of the columns too.

Adding impurities to discotic liquid crystals may also break the columns of molecules. Aliphatic molecules, when added to a columnar liquid crystal, would preferentially dissolve in the periphery of the columns because of the London interaction that makes resembling orbitals to attract one another.[14] They would first swell the aliphatic chains of the discotic molecules and increase the spacing distance between the columns.[17] At higher concentrations however, the entropy cost for arranging the solute molecules in the aliphatic region around the columns only, becomes high, and the osmotic pressure of these solute molecules gets sufficient to make inclusions into the columns and thus to break them. These point defects may then be numerous, breaking the columns in many places, and for the same entropy reasons as for the thermally produced defects, they should not coalesce in giant dislocations too. This solute effect, reducing the column length, should thus give nematic-like properties at the macroscopic scale to the columnar phase (Section 4. 3. 1). Naturally, at lower concentrations, the micro-inclusion process that breaks the columns, already works, but at a lower level. The average column length will be shorter than in the pure columnar phase ($L \sim 1.5$ mm in the pure CuD columnar phase). The mechanical properties of the strands will be modified accordingly (Section 4. 3. 5).

## 3. 1. 1. Length of the columns

So, whatever is the physical origin of the column breaks, low enough energy barrier $E_b$, or large osmotic pressure of a well-chosen solute, two extreme cases may be considered. In one case, the defects are rare, and many columns of molecules are longer than the length of the strand itself, at least for lengths below the millimetre range. Both ends of the columns then join the menisci where they may adopt complicate shapes making knots more or less difficult to untie. In this case, we anticipate that nothing will happen to the thread until the elongation stress exceeds the plasticity threshold where new point defects nucleate. Again extrapolating the results obtained in smectics, we deduce that the point defects in columnar phases are mostly repelled from the strand surface, the distortion that they produce being imaged by the free surface of the strand.[18] They therefore essentially stand around the axis of the strand. However, unlike membranes in 3D systems [19] or dislocation lines in 2D systems,[20] point defects do not exhibit any repulsive entropic interaction, because being 0D, there is only a weak chance for them to come in interaction with one another along the same axis. So, without any repulsive force, they just attract to each other as the dislocation lines do in smectics and they may eventually gather in giant dislocations.[21]

In the opposite case, when the density of point defects is large due to a low energy barrier $E_b$, or to the large osmotic pressure of an appropriate solute, the entropy cost for gathering them in giant dislocations is important compared to the potential energy earned in the aggregation process, $E_b$. The aggregation of point defects in giant dislocations should then be disfavoured so that many low order dislocations should persist in the thread. In any cases, if the elongation stress exceeds the plasticity limit, more point dislocations, or tips of columns, are created. As the initial ones, they indeed are topological defects with a +1 or -1 charge depending on whether they increase or reduce the number of columns along an oriented thread (Figure 1a). They may recombine and annihilate if they have opposite signs, so that their density should actually reach an equilibrium value. They may also move viscously under the applied stress exerted when elongating the thread, thus thinning it until eventually breaking it (Section 4. 3).

Clearly, the point defects have two consequences. They reduce the length of the columns, but also, they locally destroy the hexagonal, or rectangular, packing of the columns, so that if they become numerous enough, the phase turns to nematic in a process reminiscent to the Kosterlitz – Thouless phase transition.[22] More rigorously, we may invoke again the analogy between the 1D disordered columnar phases and the 2D smectics, and use the Helfrich's demonstration that the smectic A to nematic phase transition may be understood as a defect-mediated transition.[23]

For intermediate defect densities, the phase may however keep columnar, though exhibiting nematic-like mechanical properties. This surprising case will be discussed in Section 3. 2. 2.

### 3. 1. 2. Correlation length

Experimental results on columnar discotics are scarce yet, in particular, the natural length of the molecular columns is unknown. The density of the point defects at equilibrium should be proportional to $\exp{-\dfrac{E_b}{kT}}$ in a pure compound, and consequently, the columns should have an average length

$$L \sim h \exp{\frac{E_b}{kT}}, \tag{6}$$

where $h$ is the spacing distance between molecules in the same column. Typical values for the pure CuD columnar phase are $E_b \sim 15\ kT$ and $L \sim 1.5$ mm (Section 4. 3. 2). They should correspond to the first case discussed in Section 3. 1. 1.

Clearly, the average length of the columns $L$ determines the density of the defects $n \sim 1/LS$, $S$ being the cross-section of a column of molecules. $L$ may therefore be related to the correlation length of the phase $\xi$. Since $\xi \sim (LS)^{\frac{1}{3}}$, we should thus typically have $\xi \sim 150$ nm. Though formally, the correlation length $\xi$ could be determined from X-ray scattering, its large order of magnitude compared to the usual X-ray wavelengths makes it difficult to determine experimentally. Let us remark that usually, in the phase transition theory, the correlation length $\xi$ makes the separation between the ordered and the disordered behaviours. For sizes above $\xi$, the system behaves as in the disordered phase, and below $\xi$, it behaves as in the ordered one. Here, as the shape of the system is strongly elongated, the pertinent size that has to be compared to the strand length, is $L$. So, on strand lengths longer than $L$, the mechanical behaviour should be that of the "disordered" phase, i. e. of the nematic phase (see Section 2), and on strand lengths shorter than $L$, the behaviour should be that of the "ordered" phase, i. e. columnar. Let us finally notice that since the columns, not being perfectly oriented parallel to one another, may entangle and more or less bind together, their mechanical or effective length is indeed larger than the geometric or real length of the columns.

### 3. 1. 3. Strand of filaments

So, the disordered columnar phase is both a 1D liquid along the column axis and a 2D crystal transversally. The columns are organized in a hexagonal or rectangular array (when the molecules are tilted in the columns), so that they cannot flow freely in the perpendicular direction. The thread itself is generally not a monocrystal, except when very thin, but a polycrystal made of several crystals exhibiting slight misalignments referred to the general thread axis. They have been first observed with X-ray scattering and reported in terms of mosaicity.[24] Of diameter around $1 - 3$ µm (Section 4. 3. 2), these 2D crystals are indeed close-packed bundles of columns. They may be understood as fibrils, whiskers or filaments forming a strand. They are connected together by means of grain boundaries. The reorganization of the whole strand in a large 2D monocrystal is indeed difficult. The process may involve column realignments on propagating kinks from one end to the other. In general, this process costs a larger energy than $kT$ and does not occur spontaneously nor under the action of the surface tension alone. If not forbidden thanks to the numerous available defects (Section 3. 1. 1), the healing process may take a long time.[24] This explains that the reorganization of the strands into monocrystals does not occur practically, and in particular, that the surface tension is unable by itself to erase the grooves that the filaments form along the strand surface.

Being misaligned, the monocrystals in a discotic strand, or filaments, locally work as randomly twisted filaments. They are thus more or less entangled together.[7] Moreover, in the grain boundaries that join neighbouring filaments, individual columns, or bundles of them, may continuously run from one filament to a neighbouring one. Bridging them, they contribute also to tether the filaments together, and to give more rigidity to the strand. On the whole, both these tethering effects should provide the mechanical properties of strands made of longer columns than they are indeed. In the following, we discuss the mechanical properties of the strands in terms of the effective length of columns that is obtained on taking the tethering effects into account.

So, the geometric length of the columns is not the sole parameter to be considered for determining the mechanical properties of a strand. The statistics of the column entanglements is important too. Clearly, the entanglements of the columns should be reduced if their persistence length is much larger than their length. This may occur when the bend elastic constant of the columns is large. They should then be rather well aligned and form neatly combed filaments with only few columns bridging them together. Consequently, they will be less able to resist a mechanical stress than if dispersed and entangled over the whole strand, e. g. in the case of a smaller bend elastic constant. Cracks may then produce on breaking the fragile and few entangled columns, if they cannot directly slip out and extricate themselves. The cracks thus isolate the filaments apart from their neighbours, and allow them to slide by blocks (Figure 1b). If an appropriate solute, working as a lubricant, is then added for comparison to the discotic compound, the sliding of columns and filaments from one another, becomes even easier (Section 4. 3. 5).

## 3. 2. Elongation scenarios

Gently pulling a strand makes it to elongate, before eventually breaking down. As the volume compressibility of condensed organic matter is negligible, roughly, two extreme mechanisms may occur and possibly coexist. In the simplest case (open system), the menisci at both ends work as reservoirs to feed the strand with as much matter as necessary to preserve the integrity of the columns. The diameter is thus kept a constant all along the strand, but this model needs that the columns are long enough to be connected to a meniscus. In the other case, conversely, the volume of the strand and the number of its molecules keeps constant (closed system). This means that its matter is just redistributed inside the strand, and that its diameter and so, its number of columns, decreases during the elongation. We restrict the discussion to this latter case in the following, as it is well consistent with our observations.

### 3. 2. 1. Permeation

A simple mechanism may achieve this latter goal, namely the permeation mechanism.[4] This mechanism is favoured when the energy barrier $E_b$ for extracting or inserting a molecule from or into a column is not larger than $kT$, which leads to short and fragile columns (Section 3. 1. 1). The cohesion energy between the molecules in the columns is then limited, and it is not too difficult to shift them in the transverse direction from one column to the next one. (This makes at least two neighbouring columns to break if originally no broken columns were available in the strand, the emitting and the receiving columns). The molecules thus individually scatter among the neighbouring columns in a lateral diffusion process across the columns. This diffusion mechanism is *a priori* slow and poorly efficient, though we may suspect that it could be facilitated under an elongation stress. The molecules in an interrupted column are just submitted to the atmospheric pressure, and they keep at a normal distance to one another, while the non-broken surrounding ones undergo also an elongating stress (i. e. a negative pressure). They are consequently at a slightly larger distance than normal. It results that it is easier for them to accept new molecules from the unstressed columns, the corresponding energy barrier being lowered by the elongating stress. Most probably too, the best candidates for migrating toward surrounding columns are the molecules belonging to the very tip of broken columns, since they are poorly stabilized in comparison to well integrated discotic molecules in the rest of the column (Figure 1c). However, though these localized effects should help permeation to occur, they only concern the immediate vicinity of broken column tips. They should therefore not change much the order of magnitude of the permeation efficiency in the whole strand.

Permeation essentially being a self-diffusion process, we may appreciate its efficiency through its characteristic time for reaching equilibrium, $\tau_{perm} \sim r^2 / D$ , where $D$ is the diffusion constant of the discotic molecules across the array of columns. With $D \sim 10^{-9}$ m$^2$ s$^{-1}$ ,[25] and for a strand radius $r \sim 30$ μm, we obtain typical equilibrium times $\tau_{perm}$ of the order of a second, which seems reasonable. But, as a diffusion process, the permeation characteristic time $\tau_{perm}$ strongly depends on the radius of the strand. For radii $\sim 300$ μm, $\tau_{perm}$ would take about one minute. So, except for thin strands, or at the filament scale, the permeation process seems not efficient enough to explain the thinning of strands under an elongation stress, even if $E_b$ is smaller than $kT$ .

In the opposite case, when the energy barrier $E_b$ is large compared to $kT$, the permeation process is much less efficient again. It could however help broken columns to resorb on dispatching their molecules toward growing columns in direct contact to them (Section 3. 2. 2).

a)

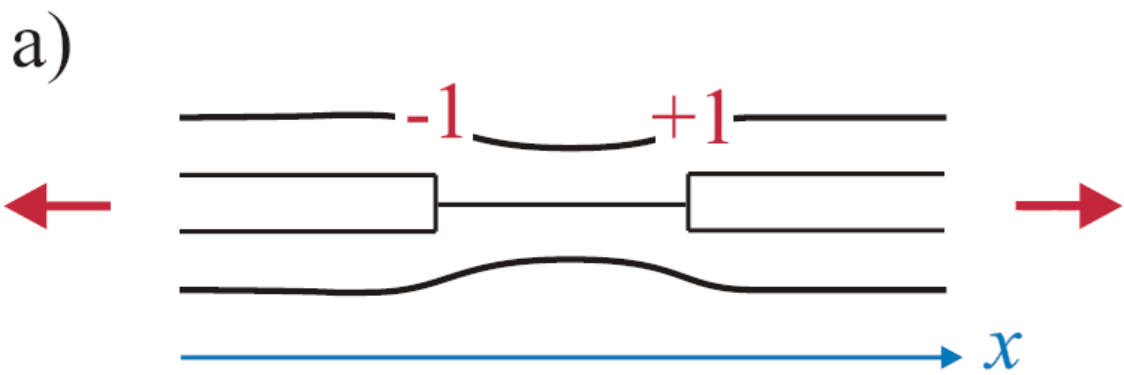

b)

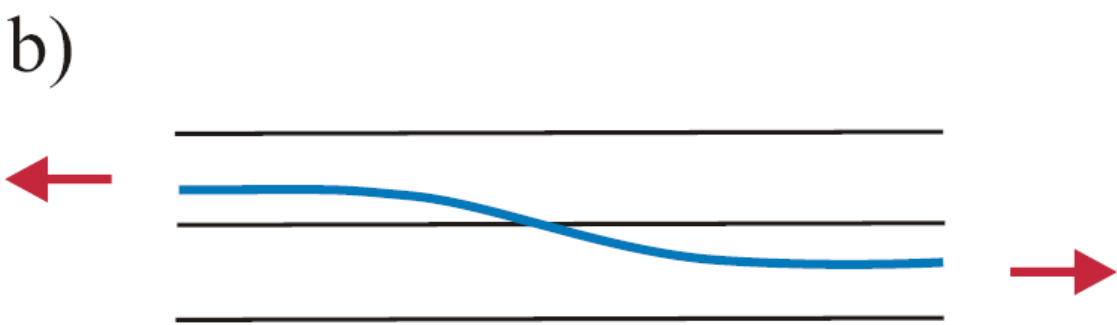

c)

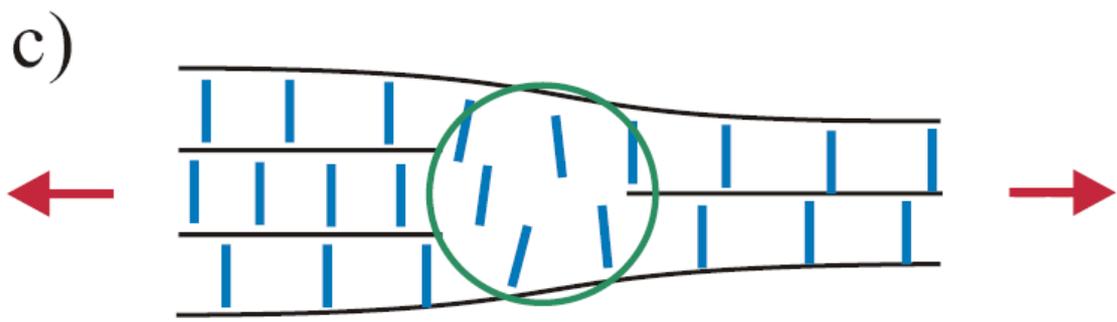

d)

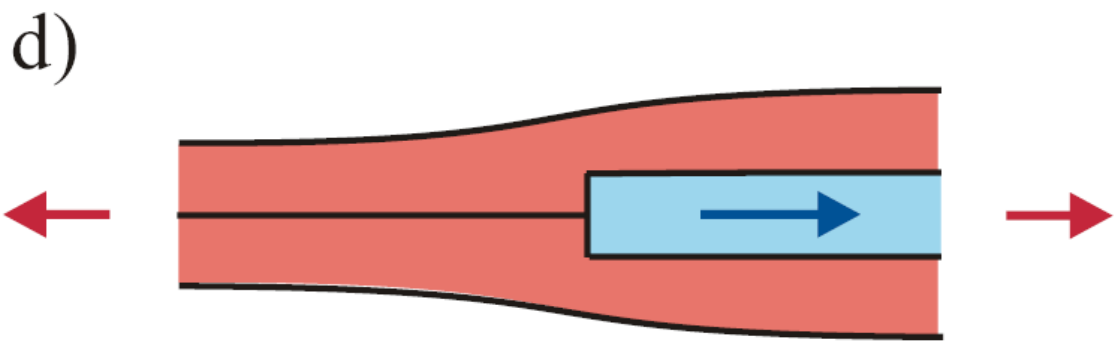

Figure 1.  Transverse sketch of a strand submitted to a stress (red arrows).

a)  Topological defects +1 or -1 increasing or decreasing the number of columns in an oriented strand, respectively.

b)  Wandering column of molecules (blue) that tethers two neighbouring filaments, or bundles of columns of molecules. Though submitted to a shear stress (red arrows), the filaments cannot slide until the tethering column is extracted out of the filaments, or broken.

c)  Discotic molecules around a column tip. They exhibit a lower positional order parameter, since on both of their sides they are in contact to contradictory columnar orders (inside the green circle). They may therefore jump more easily to neighbouring columns according to the so-called permeation process.

d)  Giant dislocation at the tip of a new filament in the strand. The dislocation produces a swelling in the strand shape that, on the photographs, looks like a ring around the strand. The filaments that are long enough to connect fixed positions on both ends, are submitted to a stress (red) until they break, while the non connected ones, if they are not tethered by wandering columns to neighbouring filaments, do not undergo any stress (blue) and may freely move away (blue arrow) all the more easily that their contact or frictional area is small.

### 3. 2. 2.  Sliding mechanism

However, in the disordered columnar phases, because the molecules in neighbouring columns have no definite positions relative to one another, no energy barrier prevents them for sliding along the column axis. The sliding of columns is then a purely viscous mechanism. Conversely, such a sliding mechanism cannot occur in ordered columnar phases, because the molecules occupy definite 3D positions that prevent them from moving. [The ordered columnar phases are indeed molecular crystals though they have unduly been classified in the liquid crystal family]. So, in the disordered columnar phases, the columns may freely slide from one another. However, as noticed in Section 3. 1. 3, such a property is not sufficient to provide fluidity to the columnar phases since the columns may be more or less entangled. The misaligned columns or bundles of columns that bridge neighbouring filaments have first to be broken before the relative sliding becomes purely viscous.

Except for this entanglement effect, the fluidity of the disordered columnar phase depends much on the density of column tips. If the broken columns are not numerous enough to provide sufficient fluidity to the strand, the stress at constant elongation velocity grows until locally exceeding the plasticity threshold, and breaks supplementary columns. In general, the column tips thus produced, preferentially locate around the strand axis, being repelled from the strand surface, and gather to form giant dislocations or cracks.[21] Nevertheless, thanks to the column compressibility, the overall profile of the strand should be significantly smoother than the shape of the dislocation itself, especially if the surface tension $\gamma$ is large compared to $\sqrt{KB}$ where $K$ is the bend elastic constant of the columns and $B$ their lateral Young modulus.[18] Due to the elongation stress, the point defects move and drive columns and bundles of them, i. e. filaments. This motion, called climb motion of the point defects in the crystallographic language, relaxes the applied stress. Indeed, the process is rather similar to tearing an ordinary strand where the individual filaments slide from one another until the strand completely breaks (Figure 1d). If the columns are properly combed in clean filaments, the two parts that are submitted to a shear stress, and that slide relative to each other, offer a lower frictional resistance to the elongation than when the columns are dispersed over the whole strand, because they present a reduced contact area. This case is opposite to the entangled $c$ase (Section 3. 1. 3). As we experimentally see below, the mechanical resistance to elongation may be changed, lowered or increased, with the addition of appropriate solutes that act as lubricants or anti-lubricants inside the contact areas between sheared filaments. The situation is then somewhat reminiscent to the use of oil (or pitch) to help (or not) drawing a rusty weapon out of its sheath.

Let us finally notice that during the elongation process, or due to the sole surface tension of the strand, the associate stress may elastically compress the columns laterally and slightly increase the spacing distance between discotic molecules. This stress is generally weak and keeps below the threshold necessary for breaking new columns. This explains why the Plateau-Rayleigh instability does not appear in columnar phases made of long columns. Conversely, if the columns have a short effective length compared to the strand length, the tethering effect between columns being taken into account (Section 3. 1. 3), nothing prevents them for sliding along the strand. Behaving as independent objects, they should globally give the mechanical properties of a 3D liquid to the strand (Section 3. 1. 1). So, the Plateau-Rayleigh instability may arise in this case of large point defect density as in a genuine nematic thread. However, the viscosity of this particular fluid should be large, essentially depending on the length of the columns that are indeed the individual objects that really move. The characteristic time before the thread ruptures may thus be long (Eq. 2). So, as noticed in Section 3. 1. 1, when the columns are much shorter than the strand itself, the strand should exhibit the mechanical properties of a nematic at macroscopic scale though the local order keeps columnar. Such a paradoxical situation seems not to be exceptional.[7] [26]

Intermediate between the disordered columnar phase made of long and resistant columns and the quasi-nematic case where the columns are short and fragile, we may naturally observe a continuous range of cases that exhibit quasi-viscous behaviours for stresses above the low plasticity threshold necessary for breaking the entangled columns. Their effective viscosity will not only depend on the shear viscosity of a perfectly oriented strand, but also on the length of the columns, their degree of entanglement, the contact areas along the fractures between sliding domains, etc… and finally may be rather large.

## 4. Experimental

We report here on experimental results obtained on pulling threads of different compounds including a polymer, in three different phases, a discotic nematic phase and two discotic disordered columnar phases, respectively, hexagonal and lamello-columnar, where the molecules are not set at definite positions in the columns. The threads are submitted to elongations by means of a mechanical set-up made of a glass needle driven by a computer-controlled stepper at a velocity that may be decreased down to a few μm per second (see Ref. [27] for details). In order to extend the study, we have doped them with two types of solute, alkyle molecules as squalane ($C_{30}H_{62}$) and triphenylene, both heavy molecules that exhibit low vapour tensions. Smaller alkyl molecules are available too, but squalane has the advantage not to evaporate significantly at our working temperature around 100°C. Adding solutes allows us to vary the phases to be studied, i.e. their nature, and their mechanical properties. We thus evidence different behaviors when pulling liquid crystal threads that, to our knowledge, have not been reported yet.

## 4. 1. Compounds

The molecules that we have studied (Figure 2) are the copper (II) dodecanoate (CuD) [28], the hexa-n-pentyloxytriphenylene (HAT5) ,[29] and the polymer PTT$_n$ (average molecular weight $M_n$

= 35.2 kDa and polydispersity index = 1.1).[30] The experiments are performed as a function of temperature and concentrations of the solutes used, alcanes or triphenylene.

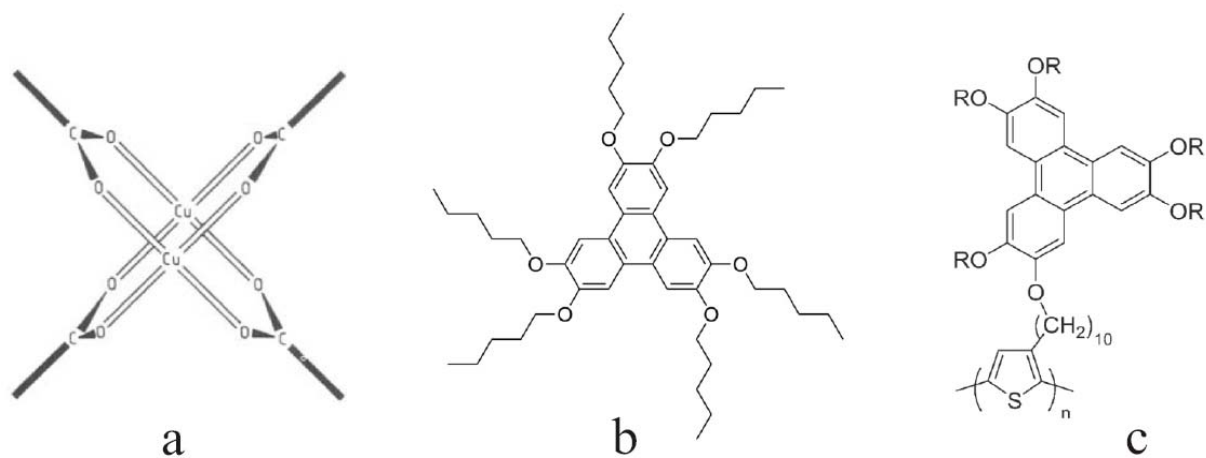

Figure 2. Chemical formulae of the compounds.

a) Copper (II) dodecanoate (CuD). The bars stand for $C_{11}H_{23}$ . Phase sequence on heating : Cryst 110°C $Col_{hex}$.

b) Hexa-n-pentyloxytriphenylene (HAT5). Phase sequence on heating : Cryst 67°C $Col_{hex}$ 121°C Iso.

c) $PTT_n$ with R = $C_5H_{11}$ : Phase sequence on heating : $Col_{rect}$ 107°C Iso ; on cooling : Iso 86°C $Col_{rect}$.

## 4. 2.  Pulling a nematic thread

The discotic threads are pulled with a glass needle in contact to a LC drop deposited on a substrate, the glass fibre being fixed to a micrometer screw driven by a computer-controlled stepper. Typically, for producing long enough nematic threads (~ 1 mm or more) to be projected on a substrate,[27] the pulling velocity has to be on the order of $v \sim 500$ µm / s. In this manner, the Plateau-Rayleigh instability has not the time to break it (Section 2. 1). The threads are prepared as symmetric as possible by means of back and forth movements until they exhibit an about cylindrical shape. They are connected on both ends to LC droplets on the substrate and on the needle by means of menisci, so that their cross-section at minimum diameter $2r$ locates around the middle. However, this shape evolves with time. Two forces amplify the thinning of the thread around the middle where it is already the thinnest, namely, the Laplace pressure and the tension $F$ exerted on the thread during its elongation. As discussed in Section 2. 2, the Laplace pressure is dominant at relatively low viscosities and elongation velocities. So, the thread middle shrinking the fastest, the thread shape becomes more or less conic on both ends after a while. The photograph in Figure 3 shows the right part of a nematic thread, with the large end connected to the meniscus (right) and the thinnest one close to the thread middle (left). The life time after the beginning of the pulling process before breaking is on the order of a couple of seconds.

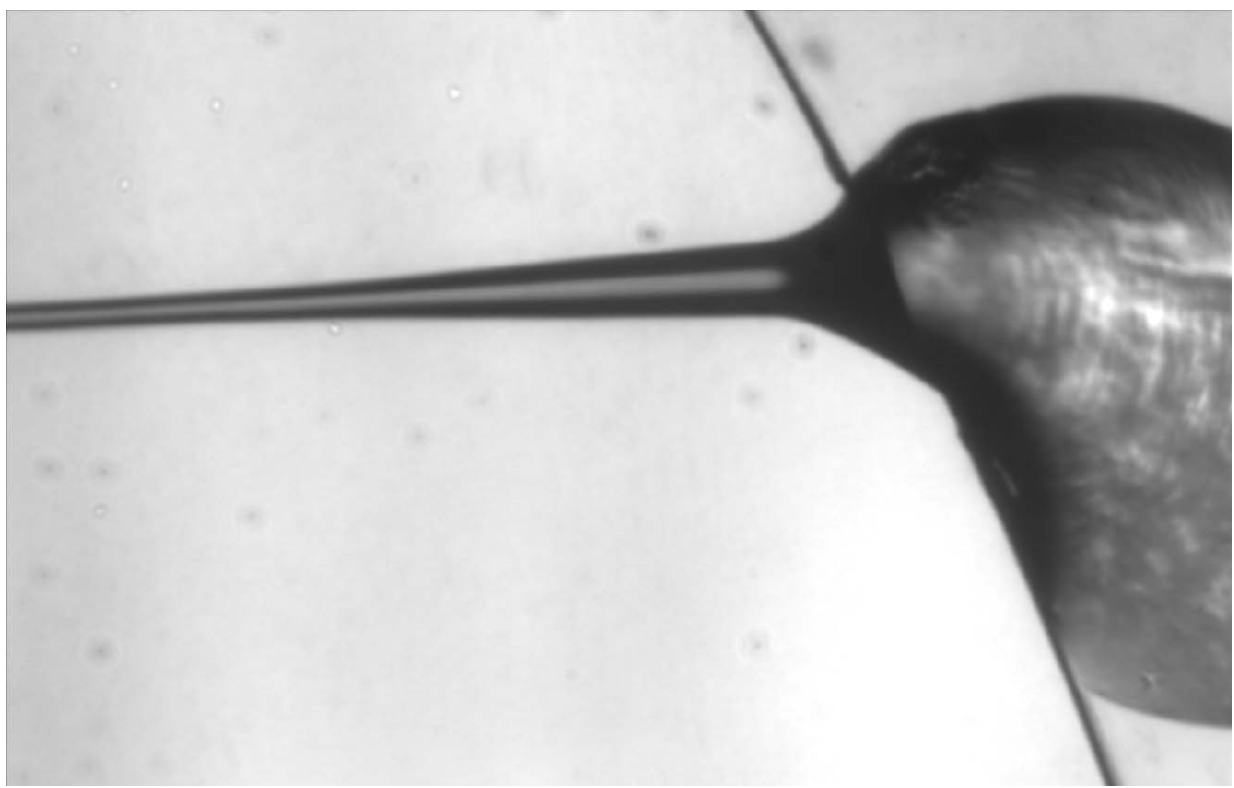

Figure 3. Free-standing thread pulled in the nematic phase at 112°C at a velocity $v = 570$ µm/s from a reservoir of a CuD - squalane mixture (80 wt% CuD, 20 wt% squalane) deposited on the edge of a cover slip (right side). In this manner, the thread – diameter of about 15 µm in its thinnest part – is stretched horizontally 100 µm above the substrate. The micrograph is shot 1.5 s after starting the pulling process, just before rupture. The length of the photo is 800 µm.

The nematic phase cannot be obtained in the pure CuD compound. Some amount of alcane molecules has to be added. Typically, we use 20 wt% of squalane, a molecule that has the advantage not to evaporate.[17] As discussed above (Eq. 4), the elongation process is essentially controlled by the ratio $\dfrac{\eta\,v}{\gamma}$. Unfortunately, the viscosity of discotic nematics is generally unknown, even more so, in the CuD - squalane mixture that we use here. We may however estimate its order of magnitude on noticing that pulling threads up to velocities $v \sim 500$ μm/s does not significantly pre-empt their rupture. We thus deduce that the Laplace pressure should not be significantly surpassed by the elongation stress up to this velocity, and that roughly then $\dfrac{l}{r} \sim \dfrac{\eta\,v}{\gamma}$. With an aspect ratio $\dfrac{l}{r} \sim 15$ (Figure 3), we deduce an estimate $\eta_{mixt} \sim 300$ Pa s that is consistent with the order of magnitude $\eta_{mixt} \sim 1000$ Pa s, obtained in Section 2. 1 from the breaking time of the thread. This is a large value, of the order of magnitude of peanut butter viscosity. An order of magnitude of the viscosity for different concentrations of the mixture may now be deduced with the Grunberg-Nissan model. On dropping the interaction terms, we may roughly estimate the viscosity of a mixture in terms of the viscosity of the virtual nematic phase in pure CuD, $\eta_{CuD}$ : [31]

$$\ln \eta_{mixt} \sim x_{CuD} \ln \eta_{CuD} + x_{Squal} \ln \eta_{Squal} \,, \tag{7}$$

where $x_{CuD} = 0.65$ and $x_{Squal} = 0.35$ are the molar fractions of CuD and squalane, respectively. Pure squalane exhibiting a viscosity $\eta_{Squal} \sim 10^{-3}$ Pa s around 100°C ,[32] we thus deduce that in the virtual nematic phase of pure CuD, $\eta_{CuD} > 1.5 \times 10^6$ Pa s. As noticed in Section 2. 2, the condition (4) is not fulfilled for so large viscosities. This means that the elongation stress is dominant in the final elongation regime, and with these values, the rupture occurs for thread diameters as large as $\sim 10$ μm (Eq. 5). In order to get thinner threads, we should decrease the pulling velocity by several orders of magnitude, which is not realistic. We may better add some solute as squalane to the CuD compound for strongly decreasing the viscosity. Then, the diameter of the thread at rupture would drastically decrease, and the rupture of the thread itself would be significantly delayed too.

So, the effects of squalane solved in discotic liquid crystals are twice. Not only, is the solute able to turn the disordered columnar phase of the CuD compound to the nematic one, but it also may significantly decrease its viscosity. The risk then is that the Plateau-Rayleigh instability significantly reduces the life time of the thread (Eq. 2). So, for realizing long and thin threads, one should choose squalane concentrations such that this life time keeps sufficiently larger than the elongation time.

## 4. 3.  Elongating disordered columnar strands

Technically, we use the same set-up to elongate strands in the disordered columnar phases as for the nematic threads, and we proceed in a similar manner.

Naturally, as recalled in Section 3, the internal structure of the columnar and smectic phases makes their elongation mechanism more complicated to discuss.[4] Dislocations in the structure are necessary for the strands to adapt their profile, at least, in their connections to the menisci. While they are edge dislocations in the case of smectics, they are point defects in the columnar phases here. This explains that to some extent both the elongations of films in the smectics and of threads in the columnar phases exhibit analogies.

Different macroscopic behaviours may be evidenced when elongating strands in the disordered columnar phases. As we discuss now, these are the strand profiles themselves that are rather irregular compared to the nematic threads, their colours between crossed polarisers, and the manner the strands evolve when elongated or conversely, when pushed back. One also may mention the absence of the Plateau-Rayleigh instability in these strands which therefore may stand for ever in the absence of any applied stress. All these features are relevant to the internal structure of the disordered columnar phase. However, in some cases, the internal structure seems to have no effect, and the strands that they produce then exhibit a surprising nematic-like behaviour.

### 4. 3. 1. Nematic-like behaviour

In Figure 4 is shown a photograph of a disordered columnar thread of pure $PTT_n$ polymer between crossed polarisers. Its optical axis is along the thread itself, indicating that the system is just uniaxial. This observation seems in contradiction with the known structure of the monomers that are strongly tilted by 39° into the columns, i. e. from the thread direction. The thread should therefore be optically biaxial.[30] In fact, this means that the thread exhibits a mosaic structure at the scale of the thread diameter ~ 10 μm, and that its optical properties are uniaxial in the average only.

Apart from slight bumps that are most probably due to inhomogeneities in the composition of the polymer, because polymers are never perfectly monodisperse, this thread seems to be rather similar to a nematic one (Figure 3). There is no macroscopic evidence of any internal structure here, though X-ray analyses have clearly shown that the structure was rectangular columnar.[30] We therefore suggest that the polymer backbone works like a solute that helps breaking the columns to much shorter lengths than the thread itself (Section 3. 1). This explains that the threads exhibit nematic-like mechanical properties though the phase is locally rectangular columnar. We may nevertheless anticipate that their effective viscosity is too large for the Plateau-Rayleigh instability breaks the thread in a practical time as for real nematic threads (Section 2. 2). The thread breaks before at finite diameters under the action of the elongation stress (Section 4. 2). The measurement of this diameter at rupture may be used to estimate the effective viscosity of the phase. Considering that the photograph in Figure 4 has been shot just before rupture, and using Eq.5, we deduce the viscosity of $PTT_n$ at 85°C to be on the order of $10^6$ Pa s.

Such a nematic-like behavior in a strand of disordered columnar strands is consistent with the mechanical observations reported by N. Clark *et al*. in HAT11,[7] and with the measurements realized by R. Stannarius *et al*. in a different compound, where in the absence of any applied stress, they have evidenced a decrease of the strand diameter over hours.[26]

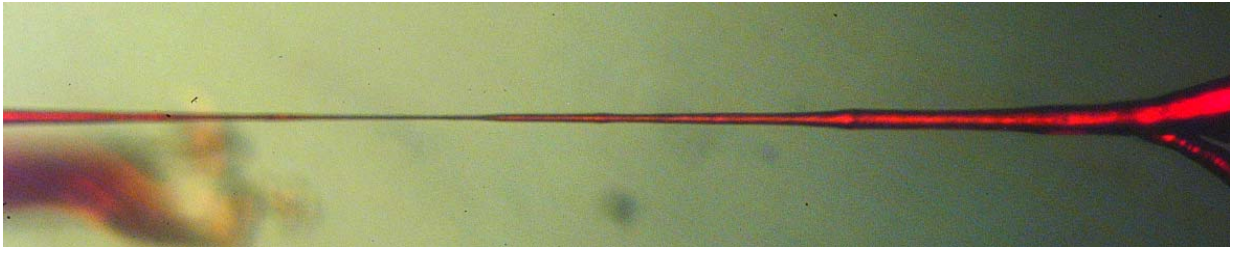

Figure 4.  Thread of pure PTT$_n$ polymer at 85°C between crossed polarisers (at 45° from the thread direction). The pulling velocity is $v = 100$ μm/s. The photo is 1.14 mm long. At minimum, the thread diameter is 3 μm. The uniform red colour is due to the absorption of the complementary green colour by PTT$_n$ . The birefringence colours usually observed in discotic strands are therefore screened out here (Section 4. 3. 3).

## 4. 3. 2.  Strand shape

At lower solute concentrations, or using less efficient solutes than the PTT$_n$ polymer backbone of the previous example to shorten the column length $L$, the disordered columnar phase should contain less point defects. The 2D crystal structure of columns is then not so finely divided as it was in the quasi-nematic case of Section 4. 3. 1. The internal structure, being globally preserved, should form small 2D crystals, or filaments, that agglomerate in strands. As discussed in Section 3, the effective length $L$ of the columns is the key parameter for the elongation mechanism of a strand, the tethering effect of misaligned columns being taken into account. We may estimate $L$ in a pure CuD columnar phase on using Eq. 6, and on taking $h \sim 0.46$ nm,[33] and $E_b = 2S\,U_{coh}$ . In this expression, the cross-section area of a column of molecules is $S \sim 3 \times 10^{-18}$ m$^2$ and $U_{coh} \sim 2 \times 10^{-2}$ J m$^{-2}$ (Section 2. 2). The cohesion energy depends both on the molecules themselves and on the possible presence of impurities. In the absence of solute, the above data yield $E_b \sim 15\,kT$ and $L \sim 1.5$ mm, and therefore $\xi \sim 165$ nm (Section 3. 1. 2). This large $E_b$ value shows that only a few point defects may spontaneously arise in the pure CuD columnar phase. However, the osmotic pressure of a solute may break columns and produce supplementary point defects (Section 3. 1. 1). As noticed above (Section 4. 2), this solute effect is strong enough to induce the nematic phase for squalane concentrations around 20 wt% and more. Naturally, for lower squalane concentrations, the micro-inclusion process already works for breaking columns, but at a lower level, with the effect of reducing the column length below the value in the pure CuD columnar phase ($L \sim 1.5$ mm).

As the sliding process reduces the elongation stress in the strand, less new columns are broken per time unit according to the plasticity process discussed above (Section 3. 2). Clearly, a competition arises between the sliding and fracture processes. The sliding may be sufficient to relax the applied stress, and therefore to stop the propagation of fractures, and to prevent new ruptures of columns. However, the efficiency of the sliding for relaxing the stress is limited since the longer the strand, the larger is the frictional area between columns of opposite sliding directions, and therefore the larger is the stress necessary for moving them. So, if the elongation process is continued at constant velocity, the stress should increase with time, until new columns break again on the most fragile places, i. e. at places where the strand is the thinnest or where broken columns have already weakened the strand. Consequently, a cyclic phenomenon takes place with the regular production of giant dislocations in the strand. Except in thin strands, they stand generally close to the strand axis, being repelled from the surface due to the surface tension $\gamma$. So, in this case of large $\gamma$ compared to $\sqrt{KB}$ (Section 3. 2. 2), the strands should exhibit rather soft profiles resembling bamboos with knots from place to place, or more exactly, resembling telescopic antennas made of tubes of decreasing diameters connected by rings (Figure 1d).

Let us now look at the general profile of elongated strands in the disordered columnar phase. In Figures 5a and 5b are shown two successive elongation states of a strand of HAT5 diluted with 1 wt % triphenylene. The images show that the strand diameter does not decrease uniformly as in the nematic phase. The strand appears to be made of segments that keep a constant diameter over a distance and that then decrease by steps on places that look like rings on the strand. These rings are consistent with the giant dislocations discussed above (Figure 1d). The analysis of the successive images of the same strand in elongation show that the thinner parts elongate more than the others and that consequently, they thin more too. Such a behaviour may be explained on noticing that the strand undergoes the largest stress where the thinnest. Therefore, the threshold for creating dislocations and cracks is overcome in the thinnest places first, and moreover, when they are created, their sliding velocity is the fastest. This amplification mechanism is indeed at the origin of the instability of the strands under elongation, which leads to their rupture. Though before elongation, the strands are prepared as cylindrical as possible on proceeding with several back and forth movements, their initial shape is always a little bit thinner around the middle between the menisci, under the action of Laplace pressure. So, when reaching their final length before rupture, i. e. about $1 - 1.5$ mm, the strands exhibit the rather conical shape shown in Figures 5a and 5b. This explains why the strands finally break at the place where they were initially the thinnest.

On carefully measuring the strand profile, we moreover observe that the thinning process works locally, without transferring matter from the menisci to the trend itself as questioned in Section 3. 2. Typically, the volume between the fixed points marked by arrows in Figures 5a and 5b keeps the same to within 10%, i. e. to within our accuracy of about 3% in the measurement of the strand sizes. This indicates that the flow of columns during the elongation takes place in both directions as well, therefore compensating any flow of matter conversely to what is proposed in Ref. [8].

Though the elongation stress at the place where the strand is the thinnest, is the largest and therefore produces the major strain, secondary stresses exist too around this spot as may be seen on comparing the strains between Figures 5a and 5b. This explains that the giant dislocations get split in several smaller ones after a while. Such a behaviour may be favoured by slight dissymmetries in the strand, and more precisely, in the tethering of its filaments to neighbouring ones by means of misaligned columns (Section 3. 1. 3). The filaments around a giant dislocation may thus be more efficiently tethered to one side than to the other one, resulting in a shear stress able to crumble the giant dislocation into several smaller dislocations. This explains that more and more rings, or bamboo knots, arise along a strand during the elongation (Figures 5a and 5b).

As noticed in Section 3. 1. 3, a strand is generally polycrystalline, and is made of imperfectly parallel filaments or fibrils (that indeed are monocrystals) stuck together. They are therefore somewhat entangled together. The diameter of these filaments depends of the cooling conditions from the nematic or from the isotropic phase. Experimentally, their diameters are found in the range $1 - 3$ μm. As whiskers in metallurgy, the filaments exhibit a crystalline structure in which any place is equivalent, so that no particular place is able to stop a fracture as soon as the rupture process is initiated. Consequently, the filaments break as a whole, and the strand gets thinner simply on reducing the number of its filaments. Naturally, this process is limited. In Figure 5c is shown a pure HAT5 strand that has been strongly elongated until reaching the monocrystal scale around its central part. It reduces to one filament then. The zoom shows the place where the last but one filament is eliminated from the strand. This filament is indeed a second crystalline domain that connects aside to the monocrystalline strand by means of a grain boundary. The places where the strand thickness reduces correspond to giant disclinations that suddenly decrease the number of filaments. They are not marked any more by rings, but by shoulders when the strand is thin enough for the giant disclination stands off-axis. The broken filament is then located to the strand side (Figure 5c, zoom). Sudden changes of birefringence colours make them easily visible (Figure 5d). On its ends, conversely, the strand gets thick and forms menisci (Figures 5b and 5c). They are reminiscent to the menisci observed in the nematic phase (Figure 3), except that they are now obtained on adding more and more giant dislocations. The superposed rings produce striations, perpendicular to the strand axis, as already observed in Ref. [8], but incorrectly interpreted in terms of buckling instability.

Grooves may also be noticed along the strands [8]. They correspond to the filaments the strands are made of. Though they somehow are reminiscent to columns of molecules, the filaments are nevertheless different and should not be confused with them especially in AFM pictures. They are indeed 2D crystals with a hexagonal frame if the discotic molecules are perpendicular to the column axis as here with the CuD and HAT5 compounds. The surface tension cannot move the columns on their frame, the realignment process needing the formation of kinks in the columns with a high cost in energy (Section 3. 1. 3). Consequently, the surface tension is unable to smooth the strand surface.

So, both the rings and the grooves are direct consequences of the 2D crystalline nature of the disordered columnar phase. Naturally, these features reveal that this phase is different from the nematic or quasi-nematic phases (Section 4. 3. 1). The differences arise essentially from the effective length of the columns $L$ compared to the length of the strand itself, the tethering effect of the columns being taken into account. The strand exhibits the mechanical properties of a columnar phase when $L$ is larger than the strand length, and of a nematic phase in the opposite case (Section 3. 1. 2). Except for these irregularities (rings and grooves) that reveal local defects in the internal structure (giant dislocations and grain boundaries, respectively), the overall profile of a strand of disordered columnar phase keeps rather similar to the nematic or quasi-nematic threads (Sections 4. 2 and 4. 3. 1, respectively).

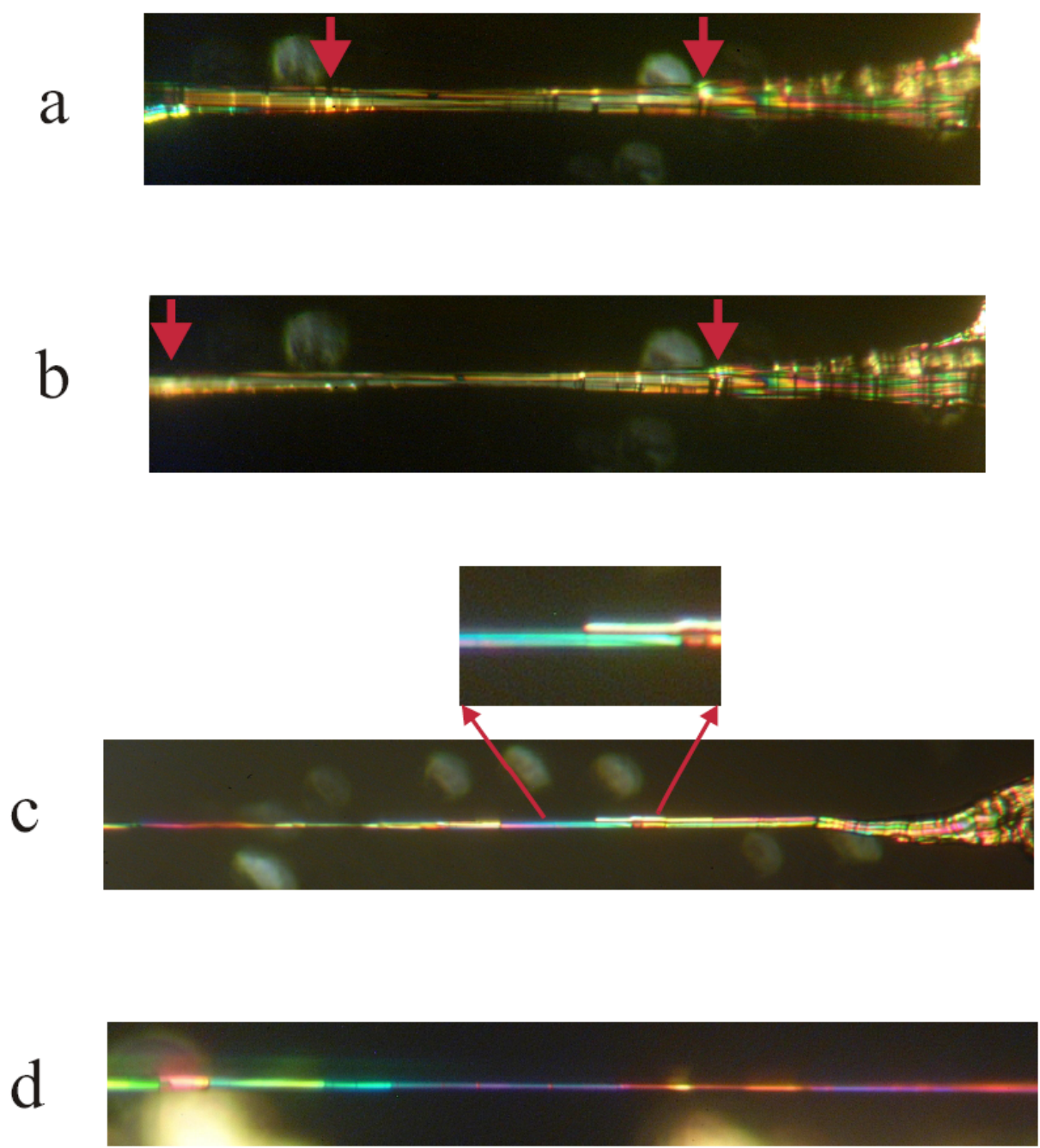

Figure 5. Microphotographs of HAT5 elongated strands between crossed polarisers.

a) and b)  Same strand of HAT5 diluted with 1 wt % triphenylene, at two successive stages of elongations. The temperature is 90°C and the elongation velocity 1.8 μm/s. The photos are 1mm long. The strand looks like a telescopic antenna. The arrows mark fixed points on the strand before and after elongation. In Figure 5a, the cylindrical part between the arrows exhibits a length and a diameter of 440 µm and 35 µm, respectively, while after additional elongation (Figure 5b), these lengths become 670 µm and 25 µm, respectively. Detailed intermediate measurements show that the volume is conserved during the elongation.

c) and d) Pure HAT5 strands pulled at the velocity of 1.8 µm/s. The polarisers are crossed at 30° from the strand axis. The temperatures of the strands in figures 5c and 5d are 90°C and 70°C, respectively. The images are 1.13 mm long. The zoom (Figure 5c) shows a second monocrystal adding up to the strand. In figure 5d, the strand diameter in the thinnest part is ~ 1.4 µm. The path difference between the ordinary and extraordinary rays is then small ~ 0.4 µm. This explains the dark violet colour there, evolving toward larger wavelength colours on the thicker sides close to this part of the strand, green on the left, red on the right. Farther again from the thinnest part of the strand, the interference order increases over 1, and therefore, the colours become difficult to interpret.

### 4. 3. 3. Birefringence colours

According to the symmetry rules, the hexagonal phase is optically uniaxial, its principal axis being oriented along the columns of molecules, i. e. along the strand axis itself. Therefore, the columnar phases exhibit optical birefringence, and in particular, the hexagonal columnar phase with the molecules being perpendicular to the columns, is uniaxial. An optical observation between crossed polarisers shows that the strands extinguish when they are oriented parallel or perpendicular to the polarisers. This simple observation confirms that the columns of molecules are essentially oriented along the strand axis.[7, 8] Now, when the polarisers are rotated at an angle, keeping perpendicular to each other, the thick strands become light, and if they are thin, they get bright uniform colours. As we may see on Figure 6, the light rays that cross over the strand, mainly follow paths of length close to, but a little less than, a diameter. This explains that the path difference between the ordinary and extraordinary rays that cross the strand, is about independent of the incident point of the light ray on the strand, and that its interferential colours are rather uniform.

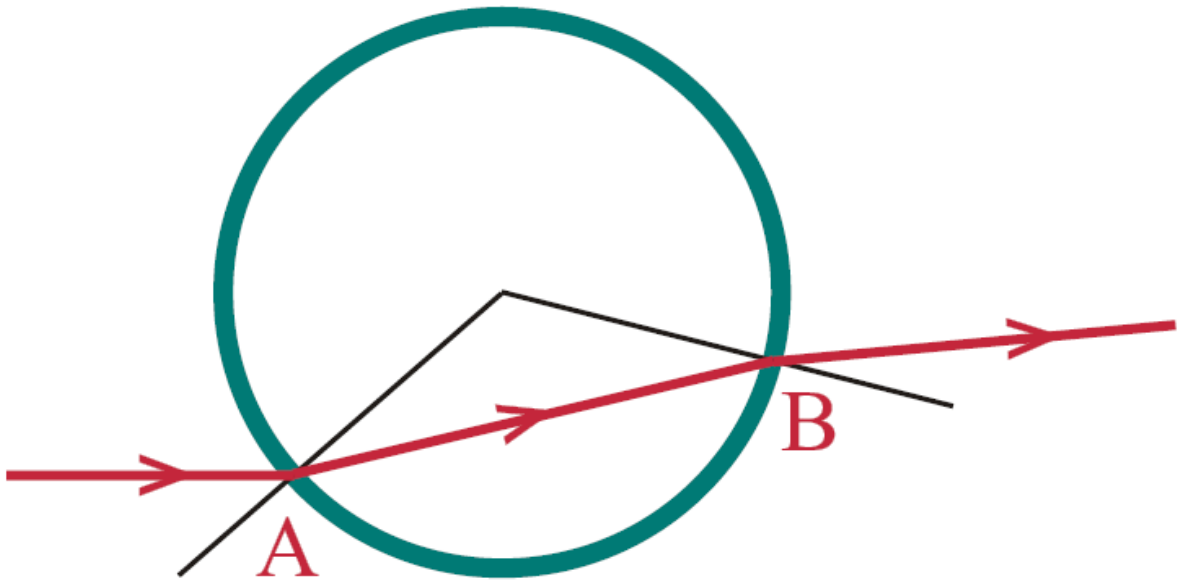

Figure 6. Cross-section of a strand. The path length AB of a light ray (red) inside a strand is close to its diameter even if a little bit off-centred. This explains that in thin strands the birefringence colours are rather independent of the distance to axis.

So, when observed between crossed polarisers, the strands in the hexagonal columnar phase of diameter in the range 1 – 10 μm generally exhibit bright colours. These colours are mainly birefringence colours due to the uniaxial symmetry of the phase. They vanish for large diameters, the interference order being too large for producing bright colours then. They are essentially relevant to the strand thickness, so that they help delimiting both the rings and the grooves discussed in Section 4. 3. 2, in terms of giant dislocations and of polycrystalline domains or filaments, respectively. As these places where the strand thickness changes, are sharp, the changes of colours are sharp too. This explains well the bamboo-like structure in Figures 5c and 5d, and the knots that sharply separate the colours.

The strands exhibit bright colours when their diameters are around 7 μm. In the thinner part (~ 1.4 μm) of Figure 5d, the colour is dark blue, indicating an interference order ~ 1 or less for the blue colour. This observation allows us to estimate the birefringence of the strand to be $\Delta n$ ~ 0.2. The value is larger than usually measured in the calamitic thermotropic liquid crystals, but this is consistent with the electron-rich core of the discotic molecules being stabilized inside the columns. In the calamitic liquid crystals, the molecules freely rotate around their axis, which produces an averaging of the transverse polarisabilities, and consequently, an increase of the optical index for the transverse polarisation and a decrease of the birefringence.[2]

The colours that we observe in the strand are not only due to thickness changes. They may also be relevant to the birefringence induced in the strand by the elongation stress, but the colours are then fainter and vary continuously along the strand. Such a stress induced birefringence is common in transparent solids, as e. g. in polymethyl methacrylate. The induced colours are reversible and disappear when the stress is suppressed on pushing the strand back.

Let us notice that the hexagonal phase is uniaxial, and that its optical properties do not depend on the exact direction of the hexagonal axes, provided they keep perpendicular to the strand axis. So, the colours of parallel filaments in a strand cannot be related to the orientation of their hexagonal axes. This would naturally not be the case for a strand in the lamello-columnar phase.

Though the colours are easy to observe, it is difficult to use them to obtain useful informations on the strand thickness or on the strand stresses. When the strand is thick, the interference order is larger than 1, and is difficult to determine exactly. When conversely the strand gets thin, its radius $r$ decreasing, its stress increases as $1/r^2$, so that the optical path difference between the ordinary and extraordinary rays in the strand increases too, proportionally to $1/r$. The stress induced birefringence therefore cannot be neglected in the thinnest strands, and its induced colours mix up with the birefringence colours, so that they are difficult to use too, for instance, for measuring the strand thicknesses accurately.

## 4. 3. 4.  Other macroscopic evidences of an internal structure

As may be seen in the right part of Figure 5c, the meniscus exhibits a hook shape. Such a shape never occurs in true liquids. This indicates that a bending moment may be sustained by the strand. This is typically a solid-like property that is relevant to the existence of an internal structure in the strand, and that confirms the previous observations performed on the strand profiles (Section 4. 3. 2). If we now push back the strand on reversing the motor rotation, we observe that the strand may bend (Figure 7b) and then coils up (Figure 7c). As the compound is non-chiral, this indicates that the chiral symmetry is broken at random on the first turn. Then, new turns grow in the same sense as the first one. When pulled again, the strand eventually breaks. We observe then that the strand keeps rigidly straight in the air, and does not fall down to the substrate as nematic threads do, because they are liquid. All these features are consistent with the structured nature of the disordered columnar phases.

a

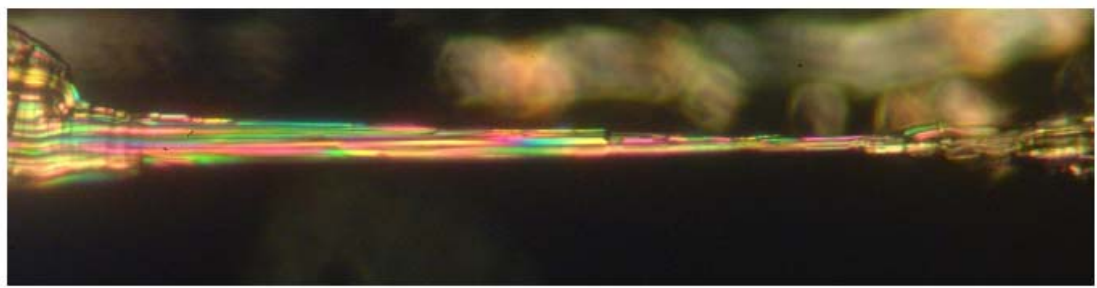

b

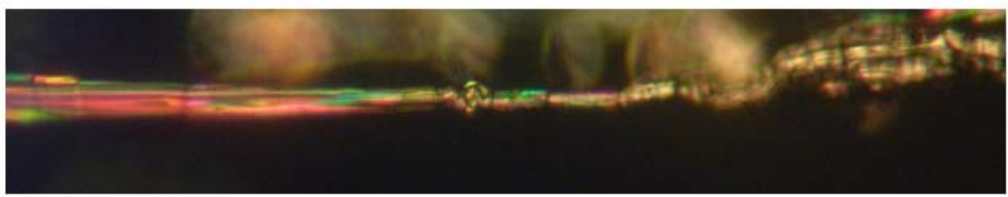

c

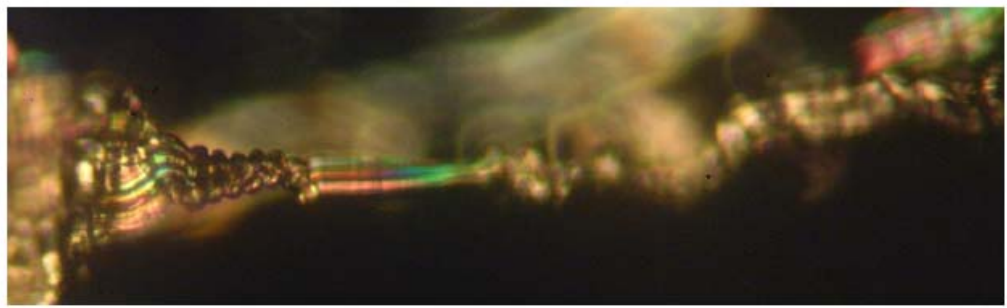

Figure 7. Same strand of HAT5 diluted with 5 wt % triphenylene. The temperature is 90°C and the elongation velocity 1.8 µm/s.

a) The image is 1.08 mm long. Being thick, the strand produces high order interferences, and therefore exhibits less bright colours than in thin strands (Figures 5c and 5d).

b) and c) Different stages of the same strand when pushed backwards. The images are 0.99 mm long. First (Figure 7b), some bends occur in some places (around the middle). The strand exhibits less colours as most of the elongation stresses are relaxed. Then (Figure 7c), the strand, though not chiral, coils up in different places.

## 4. 3. 5. Solute effects

Adding solutes to a pure compound has the well-known effect to change the temperature of the phase transitions. In some cases moreover, the nature of the phases may be modified too. For instance, if 20 wt % squalane and more, are added to the CuD compound, the disordered columnar phase turns to a nematic phase.[28] Then, pulling a thread naturally becomes easy as the nematic phase is not structured (Section 4. 2). However, even if the phase is not changed, pulling a strand is more or less difficult according to the nature of the added solute and to its concentration. Table 1 collects the diameters at rupture obtained for different strands. Two compounds, HAT5 and CuD, and two solutes, squalane and triphenylene have been tried at different concentrations. Clearly, when some amount of squalane, larger than 10 wt %, is added to CuD, the strands may get as thin as a few micrometers before breaking. Conversely, when triphenylene is added to HAT5, the reverse effect occurs, the strands breaks for larger diameters. The strands being pulled in the elongation stress regime (Section 2. 2), at the same velocity (1.8 µm/s), their minimum diameters give an estimate of their viscosity (Eq. 5).

The presence of solute molecules may modify the efficiency of permeation for transporting matter in the disordered columnar phase. However, as discussed in Section 3. 2. 1, permeation generally plays a minor role, so that we may neglect its effects when elongating strands. The dominant mechanism available to transport matter, and in this way, to change the strand shape, is the mechanism of sliding columns and filaments. In this case, the solute molecules may directly act as a lubricant. On reducing the friction between columns, they will help their sliding relative to one another. This view is consistent with the squalane molecules being strongly attracted by London interaction toward the alkyl chain region around the CuD columns. They thus preferentially swell the oily region, and increasing their thickness, they help their lubrication action.

Moreover, as noticed in Section 3. 1. 1, a second mechanism may add up for increasing fluidity. When the squalane concentration is increased, its osmotic pressure is increased too until squalane intrudes inside the columns, breaking them from place to place. Being shorter, the columns get more mobile, providing a larger fluidity to the strands. As may be seen in Table 1, both the effects help pulling CuD strands, but we cannot say yet which effect is dominant. In order to clarify this, X-ray scattering measurements of the correlation length $\xi$ as a function of the squalane concentration would be necessary. Such measurements would yield the average length of the columns $L$ in the disordered columnar phase and would show how much it depends on the squalane concentration (Section 3. 1. 2).

The case of HAT5 with triphenylene solute is particularly interesting as it exhibits a negative lubricant property of the solute. The more triphenylene is added, the more difficult are the strands to be elongated ! Such a behaviour may indeed be understood as a consequence of the London interaction that now makes the triphenylene molecules to be essentially attracted by the core of the columns, since they strongly resemble to them. They therefore swell the region of the molecule cores, and consequently, they relatively deplete the oily sheath between the columns, reducing their lubricant efficiency. As Table1 shows, the effect is sensitive. In that sense, triphenylene may be considered as an *anti*- lubricant. A few percents of triphenylene are enough to provide a much different behaviour to the strands under elongation. More precisely, as data show, the *addition* of 1wt% of triphenylene in HAT5 is roughly equivalent to a *reduction* by 2 wt% of squalane from 12 wt% to 10 wt% in CuD, as they increase the minimum strand diameter from 2 μm to 5.6 μm, and from 3 μm to9 μm, respectively. These observations are remarkably consistent, since both operations reduce the number of $CH_2$ groups in the periphery of the columns in about the same proportion of 1/30.

To our measurements, we may also add the mechanical observations reported on HAT11.[7, 24] and on $PTT_n$ strands. Compared to HAT5, HAT11 is a similar molecule, but with longer aliphatic chains. We may therefore consider that somehow the excess of aliphatic chains in HAT11 relative to HAT5 works as a solute that lubricates positively the strand and shortens its columns of molecules as squalane does in the CuD samples. This explains that longer aliphatic chains by a factor of 2, provide nematic-like properties to the strands even if the phase keeps to be disordered columnar.[7] Similarly , the aliphatic part of the $PTT_n$ backbone may be considered as a solute that lubricates positively the strand too. This explains the quasi-nematic behavior exhibited in the $PTT_n$ strands (Section 4. 3. 1).

| Compound + Solute | Strand diameter (μm) |
|---|---|
| CuD + less than 10 wt % squalane | Impossible |
| CuD+ 10 wt % squalane | 9 |
| CuD+ 12 wt % squalane | 3 |
| CuD+ 20 wt % squalane | Nematic |
| Pure HAT5 | 2.3 ; 1.5 ; 1.4 ; 2 |
| HAT5 + 1 wt % triphenylene | 5.6 |
| HAT5 + 5 wt % triphenylene | 11 |

Table 1. Strand diameters before rupture as functions of the discotic molecules and of the nature and concentration of solutes. The strands are pulled at a velocity of 1.8 μm/s.

## 5. Conclusions

In summary, the discotic liquid crystals exhibit different macroscopic behaviours when they are pulled in threads. As discussed in the paper, their mechanical properties are essentially governed by the internal structure of the phase. The simplest case is given by the nematic phase. As in any liquid, the thread roughly exhibits a figure of smooth inverted cones connected by their summit that the Plateau-Rayleigh instability breaks after a few seconds or less.

The cases of the disordered columnar phases are more complex and interesting, since according to their order at short distance, they are liquid along one dimension only. Their internal structure is then essentially characterised by the effective length of the columns of discotic molecules, which is longer than the geometric length of the columns to take into account the slight misalignments between columns and their consecutive entanglement. Roughly, there are two cases depending on whether the effective length of the columns of molecules is shorter than the length of the strand or not. In the case that the columns are shorter than the strand length, they indeed are the elementary objects of the phase that may move along the three directions of space. The phase therefore keeps fluid and surprisingly looks like a nematic phase at the macroscopic scale. In particular, the Plateau-Rayleigh instability works as in the nematic phase. However, the dynamics of this *quasi-nematic* system is much hindered compared to the true nematic case, essentially because of the large viscosity of this particular fluid made of the huge elementary objects that are the columns of discotic molecules. The effective viscosity is then so large that except for unrealistically slow pulling velocities, the elongation stress becomes dominant over the Laplace pressure in the last stages, and consequently, the strand breaks at a finite diameter (Section 2. 2). The measurement of this final diameter yields an estimate of the effective viscosity of this quasi-nematic fluid (Eq. 5).

In the opposite case of longer columns than the strand length, the Plateau-Rayleigh instability cannot arise at all, and the mechanics of the strand elongation is essentially controlled by the elongation stress. Such a behaviour clearly follows from the internal structure of the system. As shown with X-ray measurements and with the striations and grooves observed under microscope, the strands are polycrystalline. They are composed of monocrystals in the shape of filaments that indeed are bundles of columns of molecules. In general, the filaments are not perfectly parallel to one another. So, being randomly twisted from place to place, they should block themselves under traction as usual helicoidal strands do. This mechanism significantly increases the breaking strength of the whole strand.

During elongation, the strand volume keeps constant. This means that no long range transport of matter occurs, in particular, from the menisci. The strand has therefore to compensate its elongation with a simultaneous overall thinning, which may be achieved with local matter transports only. Two mechanisms are available for that. The simplest one is the permeation mechanism that corresponds to the jumping of molecules from a column to neighbouring ones. Permeation is a diffusive process, and therefore, it should essentially be efficient on short distances, typically in the case of isolated fibrils or of wandering filaments that have to get solved into the surrounding columns to improve the strand ordering. Conversely, permeation should be less efficient for relaxing stresses on large distances. Moreover, the process has to be activated to jump the energy barrier necessary to break a column, i. e. necessary to extract or to insert a molecule in a column. Permeation should therefore be less efficient in the case of compounds like pure CuD that are characterised by a large $E_b$ , i. e. that exhibit long columns of molecules.

The other available mechanism for transporting matter in a strand is based on the sliding of columns of molecules or of filaments, along their direction. However, the sliding of the columns and filaments has generally to be preceded by micro-fractures to get them free to move since they are more or less tethered with their neighbours. These fractures occur for stresses above a plasticity threshold. If a permanent stress is applied then, e. g. on continuously pulling the strand, sequential ruptures of the filaments occur and generate giant dislocations. This mechanism explains well the telescopic shapes that are observed on the elongated strands. Moreover, inhomogeneities in the entanglement of the filaments and columns may explain that shears appear on the giant dislocations when pulling the strand further, and that they finally crumble into smaller dislocations.

Finally, we have evidenced an interesting *lubricant effect* into the mechanical properties of the disordered columnar phases when some solute is added. The effect strongly depends on the nature of the solute and on its concentration. The solute may help to untie the columns from one another without breaking them, or conversely it tends to block them together. More precisely, alcanes act as lubricants, in two manners. They do not only lubricate the disordered columnar phases, whether orthogonal or tilted. They also help disorganizing the columns on inducing breaks of columns under the action of the alcane *osmotic pressure*. A quasi-nematic behaviour may then be observed when pulling strands of disordered columnar phase, but even without going so far as this rather exotic case, the viscosity of the columnar phase is substantially decreased as shown from the strand diameters before rupture. Conversely, triphenylene works as an *anti-lubricant* for triphenylene columnar systems, because it reduces the ratio of the alcane over triphenylene concentration, namely it reduces the tail over core volume ratio. The strand diameter at rupture is then observed to increase, as therefore, the viscosity.

These rather complex effects naturally open up challenging questions. Clearly, it could be of great benefit to ascertain experimentally some of the ideas discussed in the paper. An interesting point would be to determine the length of the columns of molecules. This length, that is relevant to our elongation problem, could be obtained from the correlation length $\xi$ of the disordered columnar phase. As argued in Section 3. 1. 2, $\xi$ could be in the 100 nm range, which is difficult to reach by means of neutron, or X-ray scattering, including on using synchrotron radiations. One should observe then that $\xi$ and consequently, the length of the columns depends on the discotic compound, on the added solutes and on temperature. One could then directly confirm its consequences on the detailed mechanical properties of the columnar strands under elongation, and in particular, on the strand shapes when pulling, or pushing, and on the minimum strand diameters before rupture.

Clearly too, numerical simulations could be welcome to confirm the mechanisms discussed here, as the permeation modifications around the tip of broken columns ; the permeation changes upon applied stress ; the average length of columns as a function of the energy necessary for breaking them, $E_b$ ; the role of the bend elasticity for determining the entanglement degree of the columns ; the effect of the solute concentration ; and more specifically, the lubricant or anti-lubricant effects according to the solute molecule…

## Acknowledgements

This work has been supported by the French National Agency (ANR) in the frame of its program in Nanosciences and Nanotechnologies (TRAMBIPOLY project ANR-08-NANO-051). The authors are grateful for excellent help and technical support from Nicolas Beyer and Dr. Benoît

Heinrich. The authors are also grateful to Prof. André-Jean Attias and Dr. Fabrice Mathevet for useful discussions and providing liquid crystal samples HAT5 and PTT$_n$.